\documentclass{webofc}
\usepackage[varg]{txfonts}
\usepackage{hyperref}
\usepackage{url}
\hypersetup{colorlinks=true,citecolor=blue,urlcolor=blue,linkcolor=blue}
\usepackage{mathtools}
\usepackage{amssymb}
\usepackage{physics}
\usepackage{comment}
\usepackage{xcolor}

\begin{document}

\title{p--$\phi$ femtoscopic correlation analysis using a dynamical model}
\author{\firstname{Kenshi} \lastname{Kuroki}\inst{1}\fnsep\thanks{\email{k-kuroki-e23@eagle.sophia.ac.jp}}
        \and
        \firstname{Tetsufumi} \lastname{Hirano}\inst{1}\fnsep\thanks{\email{hirano@sophia.ac.jp}}
}
\institute{Department of Physics, Sophia University, Tokyo 102-8554, Japan}
\abstract{
We analyse the p--$\phi$ correlation functions in high-multiplicity proton+proton collisions at the LHC using a dynamical model, DCCI2, and discuss the effects of collision dynamics on the p--$\phi$ femtoscopic study.
Collision dynamics, such as collective expansion and hadronic rescatterings, leads to the relative momentum-dependent non-Gaussian source functions.
This results in deviations in the correlation functions compared to those using the Gaussian source function adopted in existing studies.
Our analysis shows the importance of using the source functions that reflect more realistic collision dynamics for future precision analysis of hadron interactions via femtoscopy.
}
\maketitle

%----------------------------------------------------------
\section{Introduction}
\label{sec:intro}
Femtoscopic analysis using two-particle momentum correlations in high-energy nuclear collisions has gained significant attention as a novel phenomenological approach to understanding low-energy hadron interactions.
According to the Koonin-Pratt formula~\cite{Koonin:1977fh, Pratt:1984su}, the correlation function is interpreted as a convolution of the source function, which represents the space-time structure of the generated matter reflecting the dynamics of the nuclear collisions, and the square of the relative wave function, which reflects the final-state interaction between the pair of interest.
Thus, inputting the source function allows us to extract information about hadron interaction from the measured hadron correlation.
%普通はGaussian SFが使われていることを1行書く?
Recent active studies have demonstrated the effectiveness of femtoscopy in studying hadron interactions (see, e.g., Ref.~\cite{Fabbietti:2020bfg} for a review).
Femtoscopy is now advancing to explore less understood interactions, such as the baryon--vector meson interaction.

Recently, the ALICE Collaboration measured the proton--$\phi$ meson correlation function in high-multiplicity (0--0.17\%) proton+proton collisions at $\sqrt{s}=13~\text{TeV}$~\cite{ALICE:2021cpv}, revealing a spin-averaged attractive p--$\phi$ interaction with an unexpectedly negligible inelastic contribution.
%By fitting the data using the Lednick\'{y}-Lyuboshits model~\cite{Lednicky:1981su}, an unexpectedly sizeable scattering length with negligible inelastic contribution was extracted.
A subsequent study~\cite{Chizzali:2022pjd} reanalysed the correlation data to disentangle the two spin components of the interaction, $^2S_{1/2}$ and $^4S_{3/2}$.
By adopting the N$\phi\qty(^4S_{3/2})$ potential from the (2+1)-flavour lattice QCD calculation near the physical point~\cite{Lyu:2022imf}, the counterpart N$\phi\qty(^2S_{1/2})$ potential was constrained from the comparison with the experimental correlation function.
%The resultant scattering length of $a_0^{(1/2)}=1.54^{+0.69}_{-0.62}-i\cdot0.00^{+0.51}_{-0.00}$ indicates the controversial existing of a p--$\phi$ bound state with a binding energy in the range $\qty[12.8, 56.1]~\text{MeV}$\@.
The result indicated the controversial existence of a p--$\phi$ bound state with a binding energy in the range $\qty[12.8, 56.1]~\text{MeV}$\@.

Although simple Gaussian source functions have been assumed in existing studies, the actual hadron emission should reflect the complex dynamics of high-energy nuclear collisions.
Thus, we analyse the correlation functions using the source functions from a state-of-the-art dynamical model and discuss the effects of collision dynamics on the p--$\phi$ femtoscopy.

%----------------------------------------------------------
\section{Model and Analysis}
\label{sec:model}
As significant femtoscopic correlations %in the experimental data
appear only in the region of low relative momentum $q$, we focus exclusively on the $s$-wave scattering.
Assuming the spherical source function, the p--$\phi$ correlation function $C\qty(q)$ at the pair rest frame is expressed as
\begin{equation}
    \label{eq:KP}
    C\qty(q) = 1 + \int_0^\infty\dd{r} 4\pi r^2 S\qty(q;r) \qty{\abs{\varphi_0\qty(q;r)}^2 - \qty[j_0\qty(qr)]^2},
\end{equation}
where $r$ is the relative separation of the pair at their emission, and the source function $S\qty(q;r)$ corresponds to its probability density distribution.\footnote{
Although the relative momentum-dependence of the source function is usually neglected, the source function in principle depends on $q$.}
While $\varphi_0$ and the spherical Bessel function $j_0$ correspond to the $s$-wave parts of the radial wave functions with and without interaction, respectively, the function inside the curly bracket
%, referred to as the weight function, 
represents the increase or decrease of the wave function squared at each $r$ due to the interaction.
%According to Eq.~\eqref{eq:KP}, the correlation (i.e., the deviation of $C\qty(q)$ from unity) can be interpreted as ``how much the source function picks up the weight function''.
The wave function $\varphi_0$ can be obtained by solving the Schr\"{o}dinger equation with the p--$\phi$ interaction.

For the $^4S_{3/2}$ channel, we employ the lattice QCD potential~\cite{Lyu:2022imf}, which is overall attractive but does not support bound states.
For the $^2S_{1/2}$ channel, we follow the same parameterisation as in Ref.~\cite{Chizzali:2022pjd} motivated by the lattice QCD $V^{\qty(3/2)}$, while neglecting channel-couplings for simplicity:
\begin{equation}
    \label{eq:V}
    V^{\qty(1/2)}\qty(r)
    = \beta \qty[ a_1 e^{-\qty(r/b_1)^2} + a_2 e^{-\qty(r/b_2)^2} ] + a_3 m_\pi^4 \qty[ 1 - e^{-\qty(r/b_3)^2} ]^2 \frac{e^{-2m_\pi r}}{r^2},
\end{equation}
where $a_{1,2,3}$ and $b_{1,2,3}$ are fixed parameters listed in Table~\ref{tab:param} fitted to the lattice QCD data in the $^4S_{3/2}$ channel~\cite{Lyu:2022imf}, and they are assumed to be common to both $V^{\qty(3/2)}$ and $V^{\qty(1/2)}$.
The adjustable parameter $\beta$ is later constrained by comparison with the experimental correlation function.
Note that for $\beta=1$, the potential~\eqref{eq:V} is equivalent to $V^{\qty(3/2)}$.
%We first show the results for $\beta=7$, then vary $\beta$ and compare the resulting correlation function with the experimental data.
\begin{table}[htbp]
    \centering
    \caption{Parameter values in $V^{\qty(1/2)}$.}
    \label{tab:param}
    \begin{tabular}{ccccccc}\hline
         Parameter & $a_1~\text{[MeV]}$ & $a_2~\text{[MeV]}$ & $a_3~[\text{fm}^5]$ 
                   & $b_1~\text{[fm]}$  & $b_2~\text{[fm]}$  & $b_3~\text{[fm]}$   \\ \hline
         Value     & $-371$               & $-119$               & $-1.62$               
                   & $0.13$               & $0.3$                & $0.63$                 \\ \hline
    \end{tabular}
\end{table}

To generate the source function, we employ the dynamical core--corona initialisation model (DCCI2)~\cite{Kanakubo:2021qcw}, which is a state-of-the-art hydrodynamics-based model.
Since DCCI2 treats the QGP fluids (core) and the non-equilibrium partons (corona) simultaneously, it is well-suited for describing high-multiplicity proton+proton collisions.
From the phase space distribution of p and $\phi$ from DCCI2 at their last interacting points with the surrounding hadron gas, we obtain the p--$\phi$ source function $S\qty(q;r)$ which reflects the collision dynamics.
Note that we turn off the decay of $\phi$ to avoid the complicated reconstruction.

%----------------------------------------------------------
\section{Results}
\label{sec:results}
We impose almost the same event selection and kinematic cuts as in the ALICE measurement~\cite{ALICE:2021cpv}.
From $2.3\times10^6$ minimum bias DCCI2 simulation events for proton+proton collisions at $\sqrt{s}=13~\text{TeV}$, we select the top 0.17\% events of the charged hadron multiplicity in the pseudorapidity range $-3.7<\eta_p<-1.7$ and $2.8<\eta_p<5.1$ with at least one charged hadron in the range $\abs{\eta_p}<1$.
Then, we impose the transverse sphericity cut, $0.7<S_\text{T}<1.0$, and finally obtain 3~702 events.
In each event, protons are selected from the pseudorapidity range $\qty|\eta_p|<0.8$ and the transverse momentum range $0.5<p_\text{T}<4.05~\text{GeV}$, while $\phi$ mesons are selected from the pseudorapidity range $\qty|\eta_p|<0.8$.
Note that we employ the event-mixing method to generate the source function in the present study in order to earn statistics.
%, which guarantees the chaotic source assumption in the Koonin-Pratt formula.

\subsection{Source Function}
\label{subsec:SF}
\begin{figure}[htbp]
    \centering
    \begin{minipage}[t]{0.53\hsize}
        \includegraphics[clip,width=\linewidth]{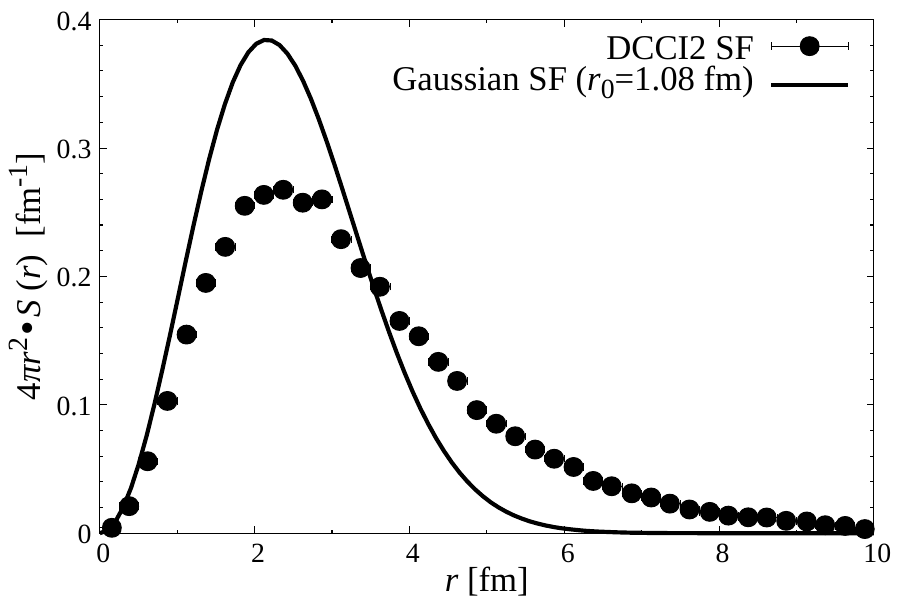}
    \end{minipage}
    \begin{minipage}[t]{0.45\hsize}
        \includegraphics[clip,width=\linewidth]{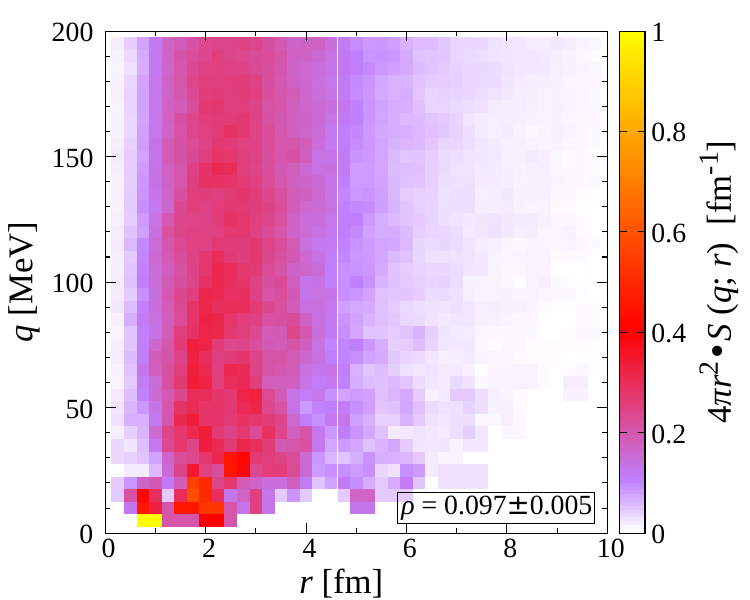}
    \end{minipage}
    \caption{Relative momentum $q$-integrated (left panel) and $q$-differential (right panel) p--$\phi$ source function multiplied by the Jacobian.
    The Gaussian source function adopted in the previous studies (solid line in the left panel) is also shown for reference.}
    \label{fig:SF}
\end{figure}
Figure~\ref{fig:SF} (left) shows the $q$-integrated source function from DCCI2 compared to the previously adopted Gaussian source function with the source size $r_0=1.08~\text{fm}$~\cite{ALICE:2021cpv, Chizzali:2022pjd}.
Due to the non-Gaussian long tail, which mainly comes from the proton rescatterings with the surrounding pion gas, the source size from DCCI2 becomes larger.

Figure~\ref{fig:SF} (right) shows the $q$-differential source function.
In general, the collision dynamics, such as the collective expansion of the generated matter, makes a correlation between the momentum space and the coordinate space, which leads to the relative momentum-dependent source function.
In fact, the source function from DCCI2 within a current kinematic setup slightly depends on $q$ with a small positive Pearson correlation coefficient $\rho\qty(q,r)\approx0.1$.
The interesting point here is that the source size in the $q\lesssim20~\text{MeV}$ region is significantly smaller than in the other regions.

\subsection{Correlation Function}
\label{subsec:CF}
\begin{figure}[htbp]
    \centering
    \sidecaption
    \includegraphics[clip,width=0.55\linewidth]{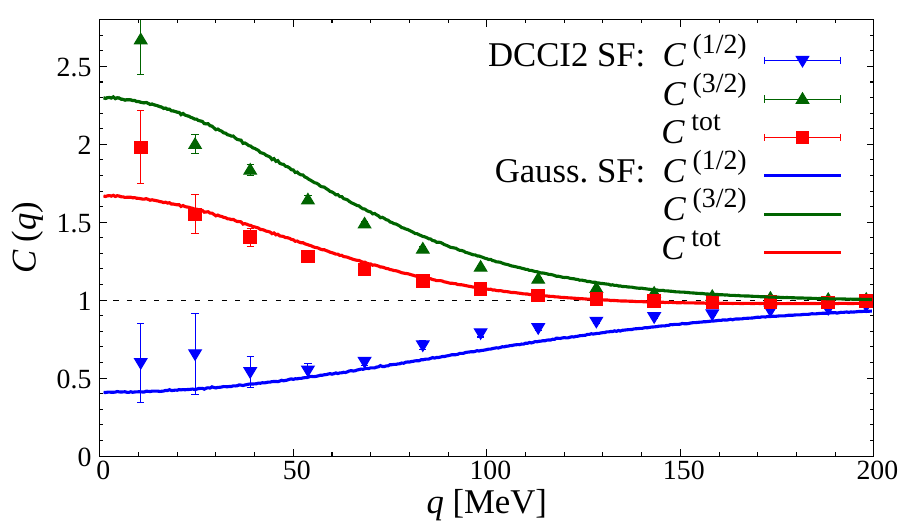}
    \caption{p--$\phi$ correlation functions using the DCCI2 source function (plots) compared with those using the Gaussian source function (lines) for the $^2S_{1/2}$ channel (blue), $^4S_{3/2}$ channel (green), and spin-average: $C^{\text{tot}}=\frac{1}{3}C^{(1/2)}+\frac{2}{3}C^{(3/2)}$ (red).
    The parameter $\beta=7$ is chosen in $V^{\qty(1/2)}$.}
    \label{fig:CF}
\end{figure}
Figure~\ref{fig:CF} shows the resulting correlation functions of each spin channel and their weighted average.
Compared to the results using the Gaussian source function, the correlation from DCCI2 is slightly weaker due to the larger source size.
In addition, one can see the intriguing behaviours in the small $q$ region due to a significantly smaller source size in this region.

Finally, we vary the adjustable parameter $\beta$ in $V^{\qty(1/2)}$ and compare the correlation function with the ALICE data~\cite{ALICE:2021cpv}.
Since $C^{(3/2)}$ is anchored by the lattice QCD potential, the spin-averaged correlation function $C^\text{tot}$, which can be compared with the experimental data, changes only through the change in $C^{(1/2)}$.
In the case of $\beta=6$, $C^{(1/2)}$ increases from unity, leading to an overestimation of $C^\text{tot}$.
On the other hand, for the cases $\beta=7$ and $8$, the overall decrease of $C^{(1/2)}$ makes $C^\text{tot}$ consistent with the ALICE data.
If we increase $\beta$ further, the suppression of $C^{(1/2)}$ becomes weaker, and thus $C^\text{tot}$ starts to overestimate the data.
From these results, $\beta$ should be around 7 to 8 within the current setup to reproduce the experimental data, which means that our femtoscopic analysis using DCCI2 also suggests the existence of a p--$\phi$ bound state in the $^2S_{1/2}$ channel with a binding energy of around 10 to 70~MeV, in agreement with a previous study~\cite{Chizzali:2022pjd}.

%----------------------------------------------------------
\section{Summary}
\label{sec:summary}
We analysed the effects of the collision dynamics on the p--$\phi$ femtoscopy in high-multiplicity proton+proton collisions at the LHC by utilising a dynamical model, DCCI2.
A non-Gaussian long tail appears in the source function mainly due to the hadronic rescatterings, leading to a slightly weaker correlation compared to that using the Gaussian source function adopted in previous studies.
In addition, we revealed that the source function slightly depends on the relative momentum due to the collision dynamics, which affects the correlation function, especially, in the small relative momentum regions.
These findings strongly suggest the importance of using the source function that reflects more realistic collision dynamics for future precision hadron interaction studies via femtoscopy.
Finally, from the comparison with the ALICE data, our analysis also supports the existence of a p--$\phi$ bound state.

%----------------------------------------------------------
\begin{acknowledgement} % KK edited line 724 in webofc.cls
The authors thank Y.~Kanakubo for generously providing the DCCI2 data essential for this study.
The work by T.~H. was partly supported by JSPS KAKENHI Grant No.~JP23K03395.
\end{acknowledgement}

%------------------------------------------------------------
\bibliography{reference} % KK edited lines 1605 and 1626 in woc.bst

\end{document}